\documentclass[aps,prd,superscriptaddress,notitlepage,nofootinbib,noeprint]{revtex4-2}
\usepackage{amsmath,amssymb,amsfonts,amsbsy}
\usepackage{xcolor}
\usepackage{graphicx}
\usepackage{multirow}
\usepackage[normalem]{ulem} 
\usepackage[makeroom]{cancel}

\begin{document}
\title{The light-by-light
contribution to the muon $g-2$ %
within the nonlocal chiral quark model with vector and axial-vector mesons}

\author{A.E. Radzhabov}
\affiliation{Matrosov Institute for System Dynamics and Control Theory SB RAS, 664033, Irkutsk, Russia}
\author{A.S. Zhevlakov}
\affiliation{Matrosov Institute for System Dynamics and Control Theory SB RAS, 664033, Irkutsk, Russia}
\affiliation{Joint Institute of Nuclear Research, BLTP,141980, Moscow region, Dubna, Russia}

\begin{abstract}
The hadronic light-by-light contribution to the muon anomalous magnetic moment is calculated in the framework of a nonlocal quark model with scalar--pseudoscalar and vector--axial-vector channels.
The effect of the dressing of internal photons by vector mesons is found to be tiny.
As a result, the contribution from the non-strange quark loop is $a_{\mu}^{\mathrm{HLbL,Loop}}=\left( 96\pm15.4\right)  \times10^{-11}$. 
The total value, including meson exchanges and the contribution of strange particles, is $a_{\mu}^{\mathrm{HLbL}}=\left( 157\pm10.6\right)\times10^{-11}$.
\end{abstract}
\maketitle
\section{Introduction}
\label{intro}
The anomalous magnetic moments of the electron \cite{Fan:2022eto,Hanneke:2008tm} and the muon \cite{Muong-2:2023cdq} are rare examples in particle physics of quantities that can be precisely measured experimentally and calculated theoretically.
To the current level of experimental precision, the anomalous magnetic moment of the electron is mostly due to quantum electrodynamics (QED) corrections with a small strong contribution\footnote{
Theoretically, the electron is known with a precision of $1.4\times10^{-11}a_e$, while the experimental precision is $1.1\times10^{-10}a_e$ \cite{Mohr:2024kco}.}
\footnote{It should be noted that there is a discrepancy between the measurement of $\alpha$ from atomic recoil measurements \cite{Parker:2018vye,Morel:2020dww} and the electron magnetic moment anomaly \cite{Fan:2022eto}. }
, while the anomalous magnetic moment of the muon is more sensitive to other interactions of the Standard Model.

In Table \ref{LBL}, the current experimental average from BNL and Fermilab \cite{Muong-2:2023cdq}, as well as the theoretical predictions from the White Paper 2020 \cite{Aoyama:2020ynm} and the SnowMass 2021 update \cite {Colangelo:2022jxc}, are presented for the different SM contributions (electromagnetic, strong, and weak)\footnote{Through the paper, the LbL contribution is given in units of $10^{-11}$.}. The difference between the experimental measurements and the theoretical predictions for the 2020-21 is a few standard deviations. This could indicate the possible influence of some unknown interaction, such as New Physics. 

However, the situation has significantly changed in recent years due to new experimental data and theoretical calculations, including lattice simulations. Experimental precision has increased, and the current world average for the muon magnetic-moment anomaly is now dominated by the Fermilab experiment. To date, Fermilab has published only the Run-2 and Run-3 datasets, and the scientific community is now awaiting the processing of new data from Run-4 and subsequent runs.

In recent years, the significant theoretical progress is related to the study of strong interaction contributions: hadronic vacuum polarization (HVP) and subleading by the fine-structure constant light-by-light scattering (LbL).
These contributions cannot be calculated theoretically using QCD perturbation theory.
They represent the dominant sources of uncertainty in the theoretical prediction of the muon $g-2$ value.
The lattice calculation from the BMW Collaboration\footnote{The uncertainties in the recent BMW result are reduced by 40 \% \cite{Boccaletti:2024guq}.} \cite{Borsanyi:2020mff} shows a sub-percent level of uncertainty for the HVP contribution, which is comparable to the precision of estimations based on $e^+e^-$-hadronic annihilation. However, there is a significant difference between these estimates. Partial estimates of the HVP contribution from other lattice groups agree with the results of calculations from collaboration BMW. The light-quark connected contribution, calculated on the basis of $e^+e^-$ hadronic annihilation data, differs significantly from lattice calculations \cite{Borsanyi:2020mff,Boccaletti:2024guq,Aubin:2022ino,Blum:2023qou,Wang:2022lkq,Ce:2022kxy,Lehner:2020crt,Alexandrou:2022amy,FermilabLatticeHPQCD:2023jof}. 
On the other hand, the new measurement of the $e^+e^- \to \pi^+ \pi^-$ cross section, from the threshold to 1.2 GeV, using the CMD-3 detector, also differs from all previous measurements \cite{CMD-3:2023alj}.

Earlier estimates of the contribution of each were based solely on effective models \cite{Bijnens:1995xf,Hayakawa:1996ki,Knecht:2001qf,Jegerlehner:2009ry}, but now estimates using other approaches have emerged. %
Recent new advances include lattice QCD results from several collaborations \cite{Chao:2022xzg,Blum:2023vlm,Fodor:2024jyn}, a dispersion-theory derivation \cite{Hoferichter:2024vbu}, and an AdS/QCD computation \cite{Leutgeb:2024rfs}, among others.

In this paper, the estimation of the light-by-light contribution within the framework of the quark model with a scalar--pseudoscalar and a vector--axial-vector sector is presented. The paper extends the calculations of the quark models with only scalar-pseudoscar particles \cite{Dorokhov:2015psa} and the calculation of resonance exchanges by adding spin-1 particles \cite{Radzhabov:2023odj}. A new effect that occurs in the model involving vector particles is the dressing of photons by intermediate vector meson exchanges. This effect was previously estimated within the Extended Nambu--Jona-Lasinio (ENJL) model \cite{Bijnens:1995xf} and Dyson--Schwinger equations (DSE) \cite{Goecke:2012qm}. In the ENJL case, this effect reduces the final answer by a factor of three, while in the DSE case it leads to only a few percent difference.

The paper is organized as follows. In Sec. \ref{model}, the model is discussed. In Sec. \ref{ExCurrents}, the vertices with external currents and vector meson dressing are discussed.
The diagrams of light-by-light processes and the calculation procedure are given in Sec. \ref{Diagrams}. 
Results are presented in Sec. \ref{Results}. In Sec. \ref{Comparison}, a comparison with other approaches is presented.
In appendix \ref{MqaqpVertex}
the nonlocal meson--quark--antiquark--photon(s) vertices are given. 
Local limits of calculations, which are important for cross-checking the calculation procedure, are considered appendix \ref{Crosscheck}.
\begin{table*}
\caption{
Muon anomaly, all numbers are given in $10^{-11}$. The experimental values are from \cite{Muong-2:2023cdq} while theoretical from  \cite{Aoyama:2020ynm,Colangelo:2022jxc}. %
}
\begin{tabular}{lc}
		\hline
Exp &  116 592 059(22)
\\
Theory& 116 591 810(43) \\
		\hline
QED &116 584 718.931(104)  \\
 EW & 153.6(10)  \\
 HVP & 6845(40)  \\
 HLbL  & 92(18)  \\
	\hline 
	\end{tabular}
	\label{LBL}
\end{table*}

\section{Model}
\label{model}
The present calculation is performed within the framework of the extended nonlocal chiral quark model, as described in references \cite{Dorokhov:2015psa,Radzhabov:2023odj, Dorokhov:2000gu}. The current-current interaction terms are taken into account in this model.
After the bosonization and Hubbard-Stratonovich transformations, the quark-meson effective Lagrangian of this nonlocal model takes the following form:\footnote{Details of the model are given in \cite{Dorokhov:2015psa,Radzhabov:2023odj}.}
\begin{align}
\mathcal{L}_{eff}&=
\bar{q}(x)(i \hat{\partial}_x -M_c)q(x)+\sigma_0 J_S^0(x) - \frac{1}{2 G_1}\left(\Big(P^a(x)\Big)^2+ \Big(\tilde{S}^a(x)+ \sigma_0 \delta^a_0\Big)^2\right)-\label{Bosonized}\\
&- \frac{1}{2 G_2}\left(\Big(V^{a,\mu}(x)\Big)^2 + \Big(A^{a,\mu}(x)\Big)^2\right)
+
P^a(x)J_{P}^{a}(x)+
\tilde{S}^{a}(x)J_{S}^{a}(x)+\notag\\
&+
V_\mu^a(x)J_{V}^{a,\mu}(x)+
A_\mu^a(x)J_{A}^{a,\mu}(x) ,\notag
\end{align}
where $M_c$ is the current quark mass matrix with diagonal elements $m_c$, $G_1$ and $G_2$ are  coupling constants in the pseudoscalar--scalar (P,S) and vector--axial-vector sectors (V,A), the mesonic fields denoted as $P$, $S$, $V$, $A$. $J_i$ are the nonlocal quark currents, which are taken in the form motivated by instanton liquid model (ILM)\footnote{In the ILM model, the form factor $f(x)$ is related to the profile function of the quark zero mode, but here it is taken as a Gaussian with a parameter fitted to mesonic observables.}\cite{Dorokhov:2000gu} 
\begin{align}
J_{M}^{a\{,\mu\}}(x)=\int d^{4}x_{1}d^{4}x_{2}\,f(x_{1})f(x_{2})\, \bar{q}
(x-x_{1})\,\Gamma_{M}^{a\{,\mu\}}q(x+x_{2}),\label{eq2}
\end{align}
with $M=S,P,V,A$, and matrices are $\Gamma_{{S}}^{a}=\lambda^{a}$, $\Gamma_{{P}}^{a}=i\gamma^{5}\lambda^{a}$, $\Gamma_{{V}}^{a,\mu}=\gamma^{\mu}\lambda^{a}$, $\Gamma_{A}^{a,\mu}=\gamma^{5}\gamma^{\mu}\lambda^{a}$. Since only light particles are considered %
the flavour matrices are: $\lambda^{a}\equiv\tau^{a}$, $a=0,..,3$ with $\tau^0=1$. The scalar isoscalar field is shifted  $S^0=\tilde{S}^0+\sigma^0$, in to obtain a physical scalar mesonic field with a zero vacuum expectation value.
$\sigma_0$ is the vacuum expectation value of the scalar isoscalar field, $\langle S^0 \rangle_0=\sigma_0\neq0$. As a result, the term in the Lagrangian $\sigma_0 J_S^0(x)$ leads to the appearance of a momentum-dependent mass $m(p)$, where the momentum dependence is factored out as
\begin{align}
m(p)=m_c+m_{d}f^2(p), \quad m_d=-\sigma_0,\label{Gap1}
\end{align}
and $m_{d}$(dynamical quark mass) can be self-consistently found from the "gap" equation
\begin{align}
m_{d}= G_1 \frac{8  N_c }{(2 \pi)^4} \int d^4_Ek
\frac{f^2(k^2)m(k^2)}{k^2+m^2(k^2)}. \label{Gap2}
\end{align}
The corresponding quark propagator is 
\begin{align}
\mathrm{S}(p)=(\hat{p}-m(p))^{-1}.
\end{align}

To describe mesonic bound states, it is necessary to calculate quark polarization loops
\begin{align}
	&\Pi_{M_1M_2}(p^2)=i \frac{N_c}{(2\pi)^4} \int d^4 k
	f^2(k_+^2)f^2(k_-^2)\, \mathrm{Tr}_{d,f}\left[ \mathrm{S}(k_-)\Gamma_{M_1}^{a}
	\mathrm{S}(k_+) \Gamma_{M_2}^{b} \right], \notag %
\end{align}
where $ k _ \pm = k \pm p / 2$ and the trace is taken over the Dirac and flavor matrices. For spin-1 bound states, the quark polarization loops and propagators should be split into transverse and longitudinal components 
\begin{align}
\Pi_{VV,AA}^{\alpha\beta}(p^2)&=\Pi_{VV,AA}^{\mathrm{T}}(p^2)\mathrm{P}^{\mathrm{T};\alpha\beta}_{p}+\Pi_{VV,AA}^{\mathrm{L}}(p^2)\mathrm{P}^{\mathrm{L};\alpha\beta}_p,
\end{align}
using projectors
\begin{align}
\mathrm{P}^{\mathrm{T};\alpha\beta}_p &= g^{\alpha\beta}- \frac{p^\alpha p^\beta}{p^2},\,\, \mathrm{P}^{\mathrm{L};\alpha\beta}_p = \frac{p^\alpha p^\beta}{p^2}\notag.%
\end{align}

Then, the unrenormalized and renormalized meson propagators (without $\pi-a_1$ mixing\footnote{Longitudinal components are related to spin-0. In the case of a system of pseudoscalar--axial-vector states, mixing appears \cite{Volkov:1986zb,Meissner:1987ge}  due to a quark polarization loop with pseudoscalar and axial-vector \cite{Radzhabov:2023odj}. The polarization loop  between scalar and vector states is identically zero.})  take the form
\begin{align}
\mathrm{D}_{S,P}(p^2)&=\frac{1}{-G_1^{-1}+\Pi_{SS,PP}(p^{2})}= \frac{g^2_{S,P}(p^2)}{p^2-M_{S,P}^2}, \quad \mathrm{D}^R_{S,P}(p^2)=\frac{\mathrm{D}_{S,P}(p^2)}{g^2_{S,P}(p^2)}\equiv\frac{1}{p^2-M_{S,P}^2}, \notag\\
\mathrm{D}_{V,A}^{\mathrm{T}}(p^2)&=\frac{1}{-G_{2}^{-1}+\Pi_{VV,AA}^{\mathrm{T}}(p^2)}= \frac{g^2_{V,A}(p^2)}{M_{V,A}^2-p^2}, \quad \mathrm{D}^{\mathrm{T};R}_{V,A}(p^2)=\frac{\mathrm{D}^{\mathrm{T}}_{V,A}(p^2)}{g^2_{V,A}(p^2)}\equiv\frac{1}{M_{V,A}^2-p^2} \quad .
	\label{pionpole}
\end{align}
The meson vertex functions  (without $\pi-a_1$ mixing) in momentum space are
\begin{align}
\mathbf{\Gamma}^{M ;\{\mu\}}_{p_1,p_2} &=g_{M}(p^2) f(p_1)\Gamma_{M}^{a\{,\mu\}}f(p_2) \label{GMesonQQ}
\end{align}
where  %
$p_i,k$ are the quark and meson momenta, respectively ($p_1=p_2+k$).

\section{External currents}
\label{ExCurrents}
\begin{figure}[t]
\centering
\begin{center}
\begin{tabular*}{0.5\textwidth}{@{}ccccc@{}}
	\raisebox{-0.5\height}{\resizebox{0.15\textwidth}{!}{\includegraphics{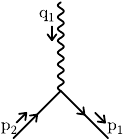}}}&&
	\raisebox{-0.5\height}{\resizebox{0.15\textwidth}{!}{\includegraphics{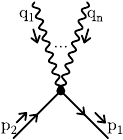}}}&&
	\raisebox{-0.5\height}{\resizebox{0.15\textwidth}{!}{\includegraphics{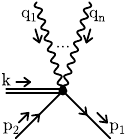}}}\\
			$(a)$&&$(b)$&&$(c)$
\end{tabular*}
\end{center}	
\caption{Photon-quark vertices: the usual local one $(a)$, the quark--antiquark vertex with n photons $(b)$ and the meson--quark--antiquark vertex with n photons $(c)$.}
\label{fig:MesonV}
\end{figure}

In the Lagrangian \eqref{Bosonized}, exist three types of terms, which in the formalism of \cite{DeWitt:1962mg,Mandelstam:1962mi} generate terms corresponding to vertices of interaction with gauge electromagnetic fields \cite{Terning:1991yt,Dorokhov:2015psa}
\begin{align}
\bar{q}(x)i \hat{\partial}_x q(x),\quad \sigma_0 J_S^0(x),\quad M^{a\{,\mu\}}(x) J_{M}^{a\{,\mu\}}(x), \label{ThreeTerms}
\end{align}
where $M$ are all possible mesonic fields. %
Therefore, the interactions with the electromagnetic field should also be included in all these terms.
In order to obtain the expressions for the vertices, the Schwinger phase factor is added to each quark field $Q(x,y)= E(x,y)q(y)$
\begin{align}
	&E(x,y)=\mathrm{exp}\left\{-i\mathrm{e}\mathrm{Q}\int \limits_x^y du_\mu G^\mu(u)\right\}, \label{Pexp}
\end{align}
where $G^\mu$ is the electromagnetic field, $\mathrm{e}$ is the elementary charge and $\mathrm{Q}$ is the charge matrix of the quark fields. Then, the following rules for the contour integral are used \cite{Terning:1991yt}
\begin{align}
	\frac{\partial}{\partial y^{\mu }}\int\limits_{x}^{y}dz_\nu G^\nu(z)=G_{\mu }(y),\quad \delta^{(4)}\left( x-y\right)
	\int\limits_{x}^{y}dz_\nu G^{\nu }(z)=0, \label{ExCuEq}
\end{align}
which guarantee the absence of non-minimal terms\footnote{It should be noted that, with these rules, the sum of the local and non-local one-photon quark-antiquark vertices leads to a result that coincides with the Ball-Chiu ans{\"a}tze \cite{Ball:1980ay}.}. The exponent inside the Schwinger factor, after expansion, generates vertices with an arbitrary number of external fields. In Fig. \ref{fig:MesonV}, the graphical representation of the vertices in \eqref{ThreeTerms} is shown: the first term corresponds to the local vertex $\mathrm{e}\mathrm{Q} \gamma^\mu$ generated by the quark kinetic term. The second term corresponds to nonlocal quark--antiquark--photon(s) vertices, and the third one corresponds to  nonlocal meson--quark--antiquark--photon(s). The expression for quark--antiquark--photon(s) vertices ${\Gamma}^{\gamma..\gamma;\mu_1..\mu_n}_{p_2,p_1,q_1..,q_n}$, with up to four photons, is given in \cite{Dorokhov:2015psa}. For meson--quark--antiquark--photon(s) ${\Gamma}^{M\gamma..\gamma;\mu_1..\mu_n}_{p_2,p_1,q_1..,q_n}$, with up to three photons, the expression is given in the appendix.

\begin{figure}[t]
\centering
\begin{center}
\begin{tabular*}{0.8\textwidth}{@{}ccccccccccccc@{}}
	\raisebox{-0.5\height}{\resizebox{0.185\textwidth}{!}{\includegraphics{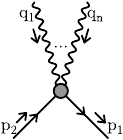}}}&{ }={ } &
	\raisebox{-0.5\height}{\resizebox{0.15\textwidth}{!}{\includegraphics{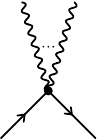}}}&{ }+{ } &
	\raisebox{-0.5\height}{\resizebox{0.15\textwidth}{!}{\includegraphics{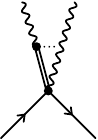}}}&{ }+{ } &{ }...{ } &{ }+{ } &
	\raisebox{-0.5\height}{\resizebox{0.15\textwidth}{!}{\includegraphics{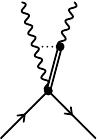}}}\\
	$(a)$&&$(b)$&&$(c)$&&&&$(d)$
\end{tabular*}
\end{center}	
\caption{The full quark--antiquark--(n)photon vertices:  with nonlocal vertex $(b)$ and with vector meson--photon transition $(c)$, $(d)$.}
\label{fig:GGaGaqq}
\end{figure}

In the presence of a vector sector, the photon(s)-quark interaction vertices are additionally dressed by a  $\rho(\omega)\rightarrow \gamma$ transition \cite{Plant:1997jr,Dorokhov:2003kf}:
\begin{align}
\mathbf{\Gamma}^{\gamma;\mu_1}_{p_2,p_1,q}&={\Gamma}^{\gamma;\mu}_{p_2,p_1,q} +\Delta\mathbf{\Gamma}^{\gamma;\mu_1}_{p_2,p_1,q_1} , \notag\\
\mathbf{\Gamma}^{\gamma\gamma;\mu_1\mu_2}_{p_2,p_1,q_1,q_2}&={\Gamma}^{\gamma\gamma;\mu_1\mu_2}_{p_2,p_1,q_1,q_2} + \Delta\mathbf{\Gamma}^{\gamma\gamma;\mu_1\mu_2}_{p_2,p_1,q_1,q_2}  , \label{AdditionalDressedVertex} \\
\mathbf{\Gamma}^{\gamma..\gamma;\mu_1..\mu_n}_{p_2,p_1,q_1..,q_n}&={\Gamma}^{\gamma..\gamma;\mu_1..\mu_n}_{p_2,p_1,q_1..,q_n} + \Delta\mathbf{\Gamma}^{\gamma..\gamma;\mu_1..\mu_n}_{p_2,p_1,q_1..,q_n},\notag
\end{align}
where $\Delta$ represents the correction of the photon(s) vertex due to vector meson dressing.
It is important to note that, from the third term in Eq. \eqref{ThreeTerms}, only one vector meson can interact with a quark-antiquark pair and external electromagnetic fields at the point of interaction. This is illustrated in Fig. \ref{fig:GGaGaqq}(c,d) for the vertex of interaction with n-photons. 
To understand the vertex corrections at the diagram level, one can join the dressing function with Fig. \ref{fig:MesonV}$(c)$ to obtain the complete expression for the vertices. For the cases of one and two photons, these expressions are:
\begin{align}
&\Delta\mathbf{\Gamma}^{\gamma;\mu_1}_{p_2,p_1,q_1}        =  
\sum\limits_{V=\rho^0,\omega}\mathbf{\Gamma}^{V;\alpha}_{p_2,p_1} \mathrm{P}^{\mathrm{T};\alpha\mu_1}_{q_1} C_{\gamma V}(q_1)\notag \\
&\Delta\mathbf{\Gamma}^{\gamma\gamma;\mu_1\mu_2}_{p_2,p_1,q_1,q_2} = 
\sum\limits_{V=\rho^0,\omega}\bigg(\mathbf{\Gamma}^{V\gamma ;\alpha,\mu_2}_{p_2,p_1,q_2}\mathrm{P}^{\mathrm{T};\alpha\mu_1}_{q_1}C_{\gamma V}(q_1)
+\mathbf{\Gamma}^{V\gamma ;\alpha,\mu_1}_{p_2,p_1,q_1}\mathrm{P}^{\mathrm{T};\alpha\mu_2}_{q_2}C_{\gamma V}(q_2)\bigg)
\end{align}
while general (n)photons vertex correction is the sum 
\begin{align}
&\Delta\mathbf{\Gamma}^{\gamma..\gamma;\mu_1..\mu_n}_{p_2,p_1,q_1,..,q_n}        =  
\sum\limits_{i=1}^n\sum\limits_{V=\rho^0,\omega}
\mathrm{P}^{\mathrm{T};\alpha\mu_i}_{q_i}C_{\gamma V}(q_i)\mathbf{\Gamma}^{V\gamma..\gamma;\alpha\mu_1..\mu_{n}}_{p_2,p_1,q_1,..,q_n}\big|_{\neq i}\notag
\end{align}  
of quark-antiquark vertices with $(n-1)$photons+vector meson. Here, %
$\mathbf{\Gamma}^{V\gamma..\gamma;\alpha\mu_1..\mu_{n}}_{p_2,p_1,q_1,..,q_n}\big|_{\neq i}$
represents a vertex with $n-1$ photons+vector meson, where the photons are taken from $1$ to $i-1$ and $i+1$ to $n$. 

This dressing is transversal and can be written in the following form:
\begin{align}
\label{PROV}
&C_{\gamma V}(q^2)=\mathrm{D}^{\mathrm{T};R}_V(q^2)  i
N_{\text{c}} \frac{\mathrm{P}^{T;\mu\nu}_{q}}{3}\int\dfrac{\mathrm{d}^{4}k}{(2\pi)^{4}}\left\{  \mathrm{Tr}%
\left[  \mathrm{S}(k_{+})\mathbf{\Gamma}^{\gamma;\mu}_{k_{+},k_{-},q} \mathrm{S}(k_{-})	\mathbf{\Gamma}^{V;\nu}_{k_-,k_+}\right]
+%
 \mathrm{Tr}\left[ \mathbf{\Gamma}^{V\gamma ;\mu\nu}_{k,k,q} \mathrm{S}(k)\right]  \right\}  ,
\end{align}
where $V$ stands for $\rho^{0}$ or $\omega$ mesons. The transition has the property $C_{\gamma V}(0)=0$ \cite{Plant:1997jr}, and it does not lead to the renormalization of the photon mass or quark charge \cite{Dorokhov:2003kf}.

\section{Diagrams}
\label{Diagrams}
The LbL contribution to the anomalous magnetic moment of the muon can be calculated using the projection\footnote{ The calculation procedure is described in detail in \cite{Dorokhov:2015psa}. } \cite{Brodsky:1966mv}:
\begin{align}
a_{\mu }^{\mathrm{HLbL}}&=\frac{1}{48m_{\mu }}\mathrm{Tr}\left( (\hat{p}%
+m_{\mu })[\gamma ^{\rho },\gamma ^{\sigma }](\hat{p}+m_{\mu })\mathrm{\Pi }%
_{\rho \sigma }(p,p)\right) , \nonumber \\
& \mathrm{\Pi }_{\rho \sigma }(p^{\prime },p)=-ie^{6}\int \frac{d^{4}q_{1}}{%
	(2\pi )^{4}}\int \frac{d^{4}q_{2}}{(2\pi )^{4}}\frac{1}{%
	q_{1}^{2}q_{2}^{2}(q_{1}+q_{2}-k)^{2}}\times  \label{P4gamProject}\\
& \quad  \times \gamma ^{\mu }\frac{\hat{p}^{\prime }-\hat{q}%
	_{1}+m_{\mu }}{(p^{\prime }-q_{1})^{2}-m_{\mu }^{2}}\gamma ^{\nu }\frac{\hat{%
		p}-\hat{q}_{1}-\hat{q}_{2}+m_{\mu }}{(p-q_{1}-q_{2})^{2}-m_{\mu }^{2}}\gamma
^{\lambda }%
\frac{\partial }{\partial k^{\rho }}\mathrm{\Pi }_{\mu \nu
	\lambda \sigma }(q_{1},q_{2},k-q_{1}-q_{2}),  \nonumber
\end{align}%
where $m_{\mu }$ is the muon mass, and
the static limit $k_{\mu }\equiv (p^{\prime }-p)_{\mu }\rightarrow 0$ is implied. 
By averaging over the direction of the muon momentum, the result for $a_{\mu }^{\mathrm{HLbL}}$ becomes a three-dimensional integral with the radial integration variables $Q_{1},Q_{2}$ and the angular variable \cite{Dorokhov:2012qa,Jegerlehner:2009ry,Jegerlehner:2017gek,Colangelo:2015ama,Colangelo:2017fiz}.

\begin{figure}[t]
\centering
\begin{center}
\begin{tabular*}{0.8\textwidth}{@{}ccccccccc@{}}
6\raisebox{-0.45\height}{\resizebox{0.12\textwidth}{!}{\includegraphics{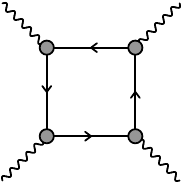}}}&{ }+{ } &
12\raisebox{-0.45\height}{\resizebox{0.12\textwidth}{!}{\includegraphics{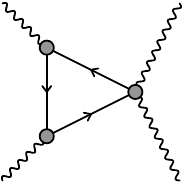}}}&{ }+{ } &
3\raisebox{-0.45\height}{\resizebox{0.12\textwidth}{!}{\includegraphics{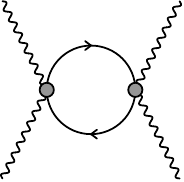}}}&{ }+{ } &
4\raisebox{-0.45\height}{\resizebox{0.12\textwidth}{!}{\includegraphics{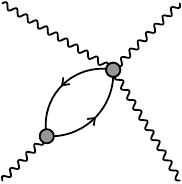}}}&{ }+{ } &
1\raisebox{-0.45\height}{\resizebox{0.12\textwidth}{!}{\includegraphics{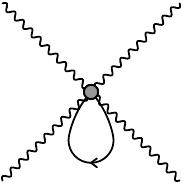}}}\\
			$(a)$&&$(b)$&&$(c)$&&$(d)$&&$(e)$
\end{tabular*}
\end{center}	
\caption{The box diagram and the diagrams with nonlocal multiphoton
interaction vertices represent the gauge invariant set of diagrams contributing to the
polarization tensor $\mathrm{\Pi}_{\mu
\nu\lambda\rho}$. The numbers in front of the diagrams are
the combinatorial factors. }%
\label{Fig:BoxCont}%
\end{figure}

The full set of diagrams for the four-rank polarization tensor $\mathrm{\Pi }_{\mu \nu\lambda \sigma }$ is similar to those in the model without the vector sector \cite{Dorokhov:2015psa}, but now all the vertices are properly dressed with vector meson exchange, as shown in Fig.\ref{Fig:BoxCont}.
As discussed in the previous section, since the model is based on the linear realization of chiral symmetry, only one $\rho(\omega)$ exchange can be present in any non-local photon-quark-antiquark vertex. 
Therefore, the exchange with one vector meson arises from diagrams Fig.\ref{Fig:BoxCont}$(a)$-$(e)$, two from Fig.\ref{Fig:BoxCont}$(a)$-$(d)$, three from Fig.\ref{Fig:BoxCont}(a)(b) and four from Fig.\ref{Fig:BoxCont}(a).
Since at zero momentum the transition loop is zero, there is no contribution $g-2$ with exchanges of $\rho(\omega)$ on the external photon line.
Therefore, the maximum number of vector meson exchanges contributing to LbL process is three.
The vector meson exchange needs to be included in all types of diagrams presented in Fig. \ref{Fig:BoxCont}$(a)$-$(e)$. External gauge field lines are free and do not include vector meson exchanges. 

\section{Results}
\label{Results}

The results in the model with spin-1 particles will differ from the result of the model without the vector meson sector for the following reasons: 
\begin{itemize}
\item the parameters of the model are different, as they are fitted to observables with the inclusion of a possible axial-vector component, 
\item there will be an intermediate axial-vector LbL contribution, 
\item the internal photon line will be dressed with intermediate vector-meson exchange.
\end{itemize}

\begin{figure}[t]
\centering
\includegraphics[width=0.49\textwidth]{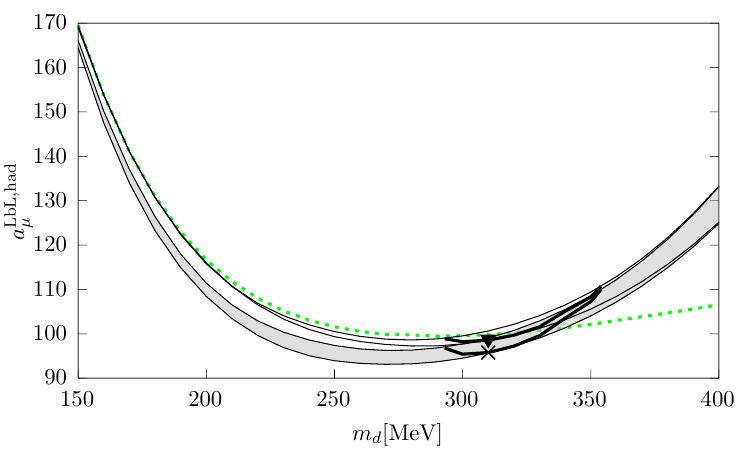}
\caption{ (in units $10^{-11}$)
Left: The LbL contribution to the muon AMM from the quark loop diagram without $\rho(\omega)$ exchanges (thin upper solid lines) and with $\rho(\omega)$ exchanges (thin downer solid lines) for different parameterizations as a function of the dynamical quark mass. The shaded area corresponds to the region between the ratios $G_1/G_2=-0.08$ and $G_1/G_2=-0.14$. The thick black solid lines correspond to the parameterization where $G_2$ is fixed in order to reproduce the physical value of the $\rho$-meson mass. The triangle corresponds to the result with $m_d=310$ MeV without $\rho(\omega)$ exchange and cross with $\rho(\omega)$.  The dashed green line is the contribution in the model without the vector sector \cite{Dorokhov:2011zf}.}
	\label{fig-QuOnlyLbl}       %
\end{figure}

\begin{figure}[t]
\centering
\includegraphics[width=0.49\textwidth]{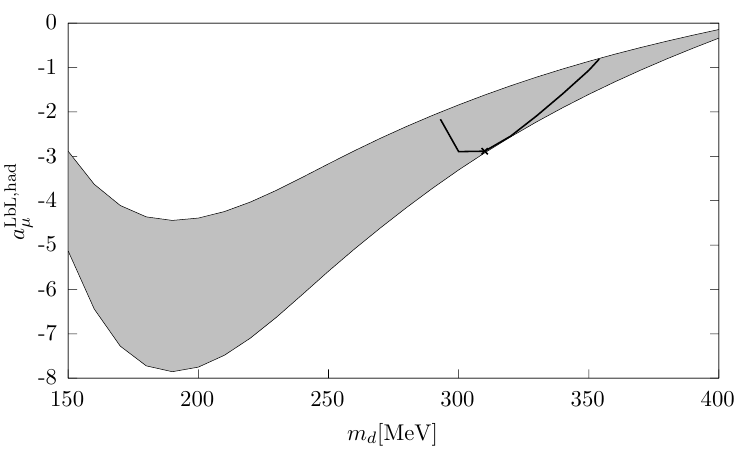}
\caption{ (in units $10^{-11}$)
Left: The LbL contribution to the muon AMM from the quark loop diagram with one intermediate $\rho(\omega)$ exchange for different parameterizations as a function of the dynamical quark mass. The shaded area corresponds to the region between the ratios $G_1/G_2=-0.08$ and $G_1/G_2=-0.14$. The black solid lines correspond to the parameterization where $G_2$ is fixed in order to reproduce the physical value of the $\rho$-meson mass. The cross corresponds to the result with $m_d=310$ MeV. }
	\label{fig-Rho1Lbl}       %
\end{figure}

\begin{figure}[t]
\centering
\includegraphics[width=0.49\textwidth]{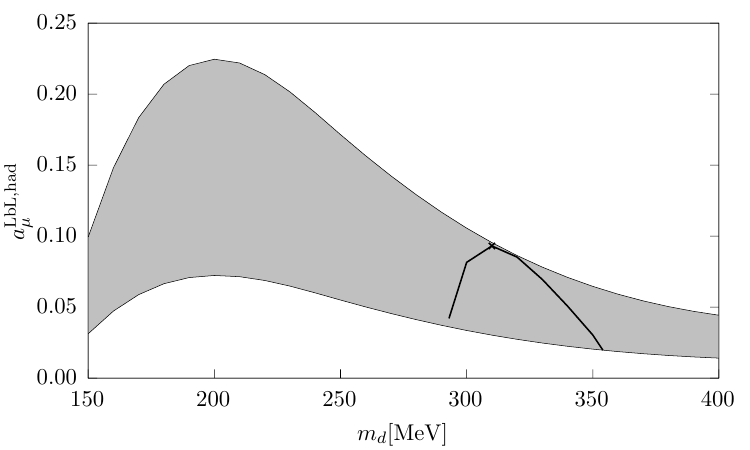}
\caption{ (in units $10^{-11}$)
Left: The LbL contribution to the muon AMM from the quark loop diagram with two intermediate $\rho(\omega)$ exchanges for different parameterizations as a function of the dynamical quark mass. The shaded area corresponds to the region between the ratios $G_1/G_2=-0.08$ and $G_1/G_2=-0.14$. The black solid lines correspond to the parameterization where $G_2$ is fixed in order to reproduce the physical value of the $\rho$-meson mass. The cross corresponds to the result with $m_d=310$ MeV. }
	\label{fig-Rho2Lbl}       %
\end{figure}

\begin{figure}[t]
\centering
\includegraphics[width=0.49\textwidth]{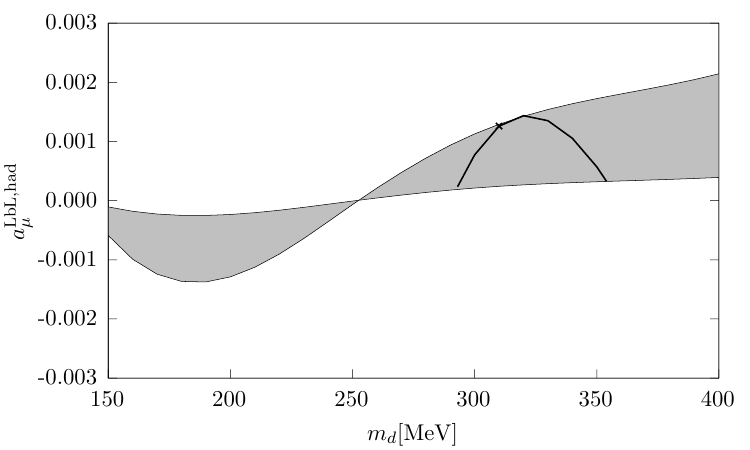}
\caption{ (in units $10^{-11}$)
Left: The LbL contribution to the muon AMM from the quark loop diagram with three intermediate $\rho(\omega)$ exchanges for different parameterizations as a function of the dynamical quark mass. The shaded area corresponds to the region between the ratios $G_1/G_2=-0.08$ and $G_1/G_2=-0.14$. The black solid lines correspond to the parameterization where $G_2$ is fixed in order to reproduce the physical value of the $\rho$-meson mass. The cross corresponds to the result with $m_d=310$ MeV. }
	\label{fig-Rho3Lbl}       %
\end{figure}

The parameters of the model are refitted in \cite{Radzhabov:2023odj} and used to calculate the LbL contribution  from  axial-vector mesons.
Due to the lack of confinement, the range of $m_d$, where the value of the $G_2$ constant can be unambiguously determined using the mass of the $\rho$-meson, is limited to between $293$ and $354$ MeV.
It is found that in this region the ratio of the four quark coupling constants, $G_2/G_1$, ranges from $-0.08$ to $-0.14$. Therefore, it is very useful to study the LbL contribution over a wider range of $m_d$ using these constant ratios.
The full quark loop contribution and the result of only diagrams without vector mesons are shown in Fig. \ref{fig-QuOnlyLbl} as a function of $m_d$. For comparison, the behavior of the quark loop contribution in a model without vector mesons is also shown \cite{Dorokhov:2015psa}. 
One can see that, for small $m_d$, the contribution from diagrams without vector mesons gives contributions similar to those in \cite{Dorokhov:2015psa}.
On the other hand, for large $m_d$, the contribution of vector mesons is very small. The individual contributions from diagrams with one, two, and three vector meson exchanges are shown in Figs. \ref{fig-Rho1Lbl}, \ref{fig-Rho2Lbl}   ,\ref{fig-Rho3Lbl}. It can be seen that the contributions of diagrams involving vector mesons are small. For the set with $m_d = 310$ MeV, these contributions are -2.9, 0.09, and 0.001, respectively.


\begin{figure}[t]
\centering
\includegraphics[width=0.49\textwidth]{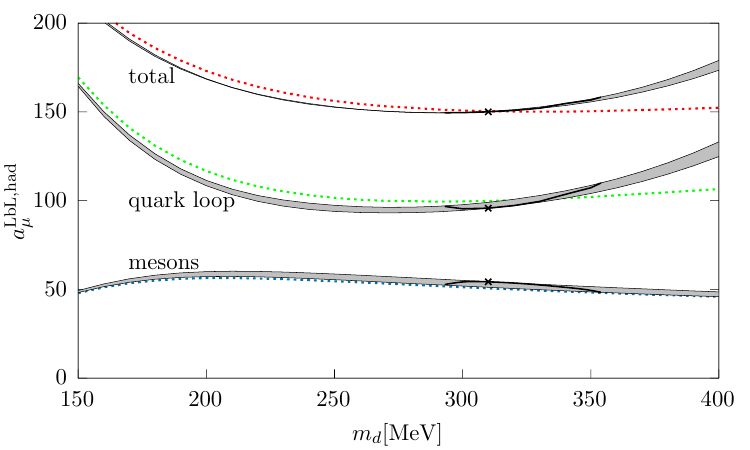}
\caption{ (in units $10^{-11}$)
Left: The LbL contribution to the muon AMM from the $\pi$, $\sigma$, $a_1$ ,$f_1$ mesons, the $u$ and $d$ quark loop, and the total $SU(2)$ contribution for different parameterizations as a function of the dynamical quark mass. The shaded area corresponds to the region between ratios $G_1/G_2=-0.08$ and $G_1/G_2=-0.14$. The thick black solid lines correspond to the parameterization where $G_2$ is fixed in order to reproduce the physical value of the $\rho$-meson mass. The cross corresponds to the result with $m_d=310$ MeV. The dashed lines are the contributions in the model without the vector sector \cite{Dorokhov:2011zf}. %
}
	\label{fig-AllQuMesonLbl}       %
\end{figure}

The combination of quark loop contributions with mesonic exchanges is given in Fig. \ref{fig-AllQuMesonLbl}. 
It is interesting to note that, for a fixed value of $m_d$, the combined contribution from quark loops and mesons is less dependent on $G_2$ compared to the individual contributions. Only for $m_d$ larger than $350$ MeV, there is still a noticeable difference between different ratios of $G_2/G_1$. Namely, for a set with $m_d = 310$ MeV, the contributions to the ratio of $-0.08$ from quark loops and mesons are $99$ and $51$, respectively, while for $-0.14$ these values are $96$ and $54$. As a result, the total value of $150$ coincides.

In the model without a vector sector \cite{Dorokhov:2015psa}, the central value is calculated as the average between the maximal and minimal values in the range $m_d$ from $200$ to $350$ MeV, while the error bar is half the difference. 
Using this strategy, the total contribution from non-strange quark loops in models with spin-1 mesons is $105\pm9.1$ which is only 3\% lower than the estimate of $108\pm8.6$ in the model without vector mesons\footnote{This value is rounded in \cite{Dorokhov:2015psa}.}. 
Adding the mesonic contributions, the $SU(2)$ result is $159\pm9.6$ with vector mesons, and $162\pm11.5$ without. 
Therefore, we can assume that the strange-particle contribution will not change much after the introduction of vector--axial-vector fields, and add to our estimate the strange-particle contribution in model without spin-1 mesons \cite{Dorokhov:2015psa}.
If we add naively the contributions of $\eta$, $\eta^\prime$, $f_0(980)$, $a_0(980)$ and the strange quark loop (since we expect that the vector meson dressing will not drastically change them) from \cite{Dorokhov:2015psa} on top of the  $SU(2)$, the resulting value becomes $166\pm10.6$, which is almost identical to the value of $168\pm12.5$ in \cite{Dorokhov:2015psa}.

However, in the model with vector mesons, it is proposed to change the method of calculating the central value.
It seems more natural to take the central value at the point where it is possible to describe the vector meson and fix the $G_2$ constant using the physical value of the $\rho$-meson mass \cite{Radzhabov:2023odj}. It is suggested to use the central value of $m_d = 310$ MeV, as at this point, it is also possible to describe the axial-vector meson.
In this case nonstrange quark loop is $96\pm9.1$, quark loop plus $SU(2)$ mesonic contribution is $150\pm9.6$. The final estimate of the contribution together with the strange particles is  %
\begin{align}
a_{\mu}^{\mathrm{HLbL}}=\left( 157\pm10.6\right)\times10^{-11}. \label{FinalResult}
\end{align}

\section{Comparison with other approaches and summary}
\label{Comparison}
\begin{figure}[t]
\centering
\includegraphics[width=0.49\textwidth]{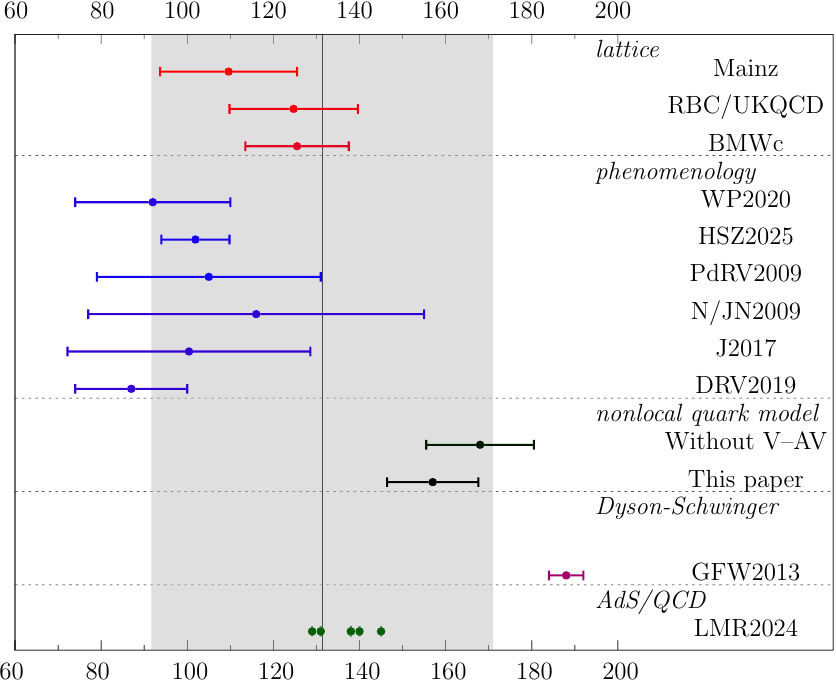}
\caption{ (in units $10^{-11}$)
The LbL contribution to the muon AMM: lattice QCD calculations  (Mainz)\cite{Chao:2022xzg}, (RBC/UKQCD)\cite{Blum:2023vlm} and (BMWc)\cite{Fodor:2024jyn}, phenomenological estimations ``White paper'' 2020 \cite{Aoyama:2020ynm}, recent evaluation using a dispersive formalism (HSZ2025)\cite{Hoferichter:2024vbu}, using combinations of different contributions 
(PdRV2009)\cite{Prades:2009tw}, (N/JN2009)\cite{Jegerlehner:2009ry}, (J2017)\cite{Jegerlehner:2017gek}, (DRV2019)\cite{Danilkin:2019mhd}, in nonlocal quark model without vector and axial-vector mesons \cite{Dorokhov:2015psa} and present paper calculations with spin-1 particles, DSE\&BSE studies (GFW2013)\cite{Goecke:2012qm}, AdS/QCD result for different models (LMR2024)\cite{Leutgeb:2024rfs}. The thin vertical and shaded region corresponds to difference between experimental value and theoretical prediction with recent lattice HVP value \cite{Boccaletti:2024guq} and without HLbL contribution. }
	\label{fig-LblAllExtended}       %
\end{figure}	

The comparison of our result \eqref{FinalResult} is given in Fig. \ref{fig-LblAllExtended} with recent lattice results by (Mainz) \cite{Chao:2022xzg}, (RBC/UKQCD) \cite{Blum:2023vlm} and (BMWc) \cite{Fodor:2024jyn}, phenomenological estimations :``White paper'' 2020 \cite{Aoyama:2020ynm}, dispersive formalism (HSZ2025)\cite{Hoferichter:2024vbu}, ``Glasgow consensus''  \cite{Prades:2009tw},   DSE\&BSE studies \cite{Goecke:2012qm}, AdS/QCD result for different models \cite{Leutgeb:2024rfs}.
In the current approach, the QCD asymptotics are automatically fulfilled by diagram with quark loops \cite{Dorokhov:2015psa} and the diagrams with vector mesons dressing of photon and resonance exchanges do not change this picture\footnote{The suppression of quark loop contributions by soft gluon fields is discussed in \cite{Nedelko:2021dsh}. }. This is similar to DSE\&BSE studies \cite{Goecke:2012qm} while in dispersive formalism \cite{Hoferichter:2024vbu} the matching to short-distance constraints\footnote{Short-distance constraints with gluonic corrections is given in \cite{Bijnens:2024jgh}. } is used. In the AdS/CFT correspondence \cite{Leutgeb:2024rfs, Leutgeb:2019gbz, Cappiello:2019hwh,}, this matching occurs through the summation of a tower of resonances.
One can see that our final result of is larger than the dispersive approach, the lattice QCD calculation and AdS/QCD.
However, it is interesting to compare how the different results in Fig. \ref{fig-LblAllExtended} relate to the difference between theoretical calculations and experimental measurements. Using the values from Table \ref{LBL}, i.e., WP2020, one can conclude that, even with an increase in the LbL contribution of $65$, there is still a gap between theory and experiment.
However, as discussed in the Introduction, there are indications from lattice calculations and CMD-3 measurements of the $e^+e^-$ annihilation cross section that the HVP value should probably be increased.
Therefore, for comparison, the difference between the experimental value and the theoretical prediction with the recent lattice HVP value \cite{Boccaletti:2024guq} and without HLbL contribution is also shown in Fig. \ref{fig-LblAllExtended}.
One can see that, although our results are larger than those obtained using the dispersive approach and lattice LbL calculations, all results based on the HVP value from \cite{Boccaletti:2024guq} suggest that the window for New Physics scenarios is becoming narrower.  

The source of discrepancy between calculations in nonlocal model and dispersive formalism is partially know.
Namely, the mean-field calculations in the quark model do not include the pion loop contribution, which is known to give a negative contribution \cite{Aoyama:2020ynm,Miramontes:2021exi}. The pion loop (or mesonic loop in general) corresponds to subleading terms in the ${1}/{N_c}$ expansion, which can be accounted for using the scheme presented in \cite{Radzhabov:2010dd}. However, in the ${1}/{N_c}$ expansion, there will be additional terms that may also affect the contact term. On the other hand, within a nonlocal model, the two-photon form factors of mesons depend not only on the photon virtualities but also on the mesonic virtuality. This additional dependence\footnote{The discussion of the dependence of the meson form factor on meson virtuality and dispersive result can be found in \cite{Hayakawa:2025mmy}.} results in a significant suppression of the light-by-light contribution from massive resonances \cite{Dorokhov:2015psa,Radzhabov:2023odj}. 

In order to better understand the difference and as an independent crosscheck, it would be interesting to repeat the calculations within the quark model framework in conditions similar to those used in lattice calculations, that is, in coordinate space with semi-analytical QED kernel \cite{Asmussen:2022oql}. This would allow us to compare not only the total number, but also the contributions from different distances, as shown in Fig.3 of Ref.\cite{Chao:2021tvp}.

The results obtained within the framework of the nonlocal quark model for the LbL contribution to the muon magnetic-moment anomaly are shown to be stable when extending from the S-PS case to include V-AV currents\footnote{A similar effect is seen in the NJL model describing $\pi\pi$ scattering \cite{Bernard:1995hm}.}. The next step is to estimate the pion loop contribution, which is a 1/$N_c$ correction and could play an important role in the overall estimation of the LbL value.

Diagrams are drawn with the help of the \verb|feyn.gle| package \cite{Grozin:2022fde}.

The authors thank the A.P. Martynenko, F.A. Martynenko and V.P. Lomov for fruitful comments.

This work is supported by the Russian Science Foundation (Grant No. RSF 23-22-00041).

\appendix
\section{Meson--quark--antiquark--photon(s) vertices}
\label{MqaqpVertex}
Due to the nonlocal interaction between mesons and quarks, vertices with one, two, or three photons, using rules \ref{ExCuEq}, take the form ($p_1=p_2+q_1+...+q_i$)
\begin{align}
\mathbf{\Gamma}^{M\gamma ;\{\alpha\}\mu_1}_{p_1,p_2,q_1}&= -\mathrm{e}g_{M}(p^2)\bigg(J_1^{\mu_1}(p_1,-q_1)\mathrm{Q}\Gamma_{M}^{a\{,\alpha\}}f(p_2)
        +f(p_1) \Gamma_{M}^{a\{,\alpha\}}\mathrm{Q} J_1^{\mu_1}(p_2,q_1)\bigg)\\
\mathbf{\Gamma}^{M\gamma\gamma ;\{\alpha\}\mu_1\mu_2}_{p_1,p_2,q_1,q_2} &=\mathrm{e}^2 g_{M}(p^2)\bigg(J_2^{\mu_1\mu_2}(p_1,-q_1,-q_2)\mathrm{Q}^2\Gamma_{M}^{a\{,\alpha\}}f(p_2)
   +J_1^{\mu_1}(p_1,-q_1)\mathrm{Q}\Gamma_{M}^{a\{,\alpha\}}\mathrm{Q}J_1^{\mu_2}(p_2,q_2) \notag\\&
     \quad+J_1^{\mu_2}(p_1,-q_2)\mathrm{Q}\Gamma_{M}^{a\{,\alpha\}}\mathrm{Q}J_1^{\mu_1}(p_2,q_1) 
   +f(p_1)\Gamma_{M}^{a\{,\alpha\}}\mathrm{Q}^2J_2^{\mu_1\mu_2}(p_2,q_1,q_2)\bigg)\\
\mathbf{\Gamma}^{M\gamma \gamma\gamma;\{\alpha\}\mu_1\mu_2\mu_3}_{p_1,p_2,q_1,q_2,q_3} &= -\mathrm{e}^3 g_{M}(p^2)\bigg(J_3^{\mu_1\mu_2\mu_3}(p_1,-q_1,-q_2,-q_3)\mathrm{Q}^3  \Gamma_{M}^{a\{,\alpha\}} f(p_2)\notag\\&
  \quad+J_2^{\mu_1\mu_2}(p_1,-q_1,-q_2)\mathrm{Q}^2\Gamma_{M}^{a\{,\alpha\}}\mathrm{Q} J_1^{\mu_3}(p_2,q_3)
  +J_2^{\mu_1\mu_3}(p_1,-q_1,-q_3)\mathrm{Q}^2\Gamma_{M}^{a\{,\alpha\}}\mathrm{Q} J_1^{\mu_2}(p_2,q_2)\notag\\&
  \quad+J_2^{\mu_2\mu_3}(p_1,-q_2,-q_3)\mathrm{Q}^2\Gamma_{M}^{a\{,\alpha\}}\mathrm{Q} J_1^{\mu_1}(p_2,q_1)
  +J_1^{\mu_3}(p_1,-q_3)\mathrm{Q}\Gamma_{M}^{a\{,\alpha\}}\mathrm{Q}^2J_2^{\mu_1\mu_2}(p_2,q_1,q_2)   \notag\\&   
  \quad+J_1^{\mu_1}(p_1,-q_1)\mathrm{Q}\Gamma_{M}^{a\{,\alpha\}}\mathrm{Q}^2J_2^{\mu_2\mu_3}(p_2,q_2,q_3)        
  +J_1^{\mu_2}(p_1,-q_2)\mathrm{Q}\Gamma_{M}^{a\{,\alpha\}}\mathrm{Q}^2J_2^{\mu_1\mu_3}(p_2,q_1,q_3)   \notag\\&     
  \quad+f(p_1)\Gamma_{M}^{a\{,\alpha\}}\mathrm{Q}^3J_3^{\mu_1\mu_2\mu_3}(p_2,q_1,q_2,q_3)\bigg)
\end{align}
where 

\begin{align}
J_1^{\mu_1}(k,q_1)&= (k+k_{1})^{\mu_1}\mathrm{f}^{(1)}_{k,k_{1}} \\
J_2^{\mu_1\mu_2}(k,q_1,q_2) &=
+(k+k_{1})^{\mu_1}(k_{1}+k_{12})^{\mu_1}\mathrm{f}^{(2)}_{k,k_{1},k_{12}} \notag\\& 
+(k+k_{2})^{\mu_2}(k_{2}+k_{12})^{\mu_2}\mathrm{f}^{(2)}_{k,k_{2},k_{12}}
+2g^{\mu_1\mu_2}\mathrm{f}^{(1)}_{k,k_{12}} \\
J_3^{\mu_1\mu_2\mu_3}(k,q_1,q_2,q_3) &=
+(k+k_{1})^{\mu_1} (k_{1}+k_{12})^{\mu_2} (k_{12}+k_{123})^{\mu_3}\mathrm{f}^{(3)}_{k,k_{1},k_{12},k_{123}}\notag\\& 
+(k+k_{1})^{\mu_1} (k_{1}+k_{13})^{\mu_3} (k_{13}+k_{123})^{\mu_2}\mathrm{f}^{(3)}_{k,k_{1},k_{13},k_{123}}\notag\\& 
+(k+k_{2})^{\mu_2} (k_{2}+k_{12})^{\mu_1} (k_{12}+k_{123})^{\mu_3}\mathrm{f}^{(3)}_{k,k_{2},k_{12},k_{123}}\notag\\& 
+(k+k_{2})^{\mu_2} (k_{2}+k_{23})^{\mu_3} (k_{23}+k_{123})^{\mu_1}\mathrm{f}^{(3)}_{k,k_{2},k_{23},k_{123}}\notag\\& 
+(k+k_{3})^{\mu_3} (k_{3}+k_{13})^{\mu_1} (k_{13}+k_{123})^{\mu_2}\mathrm{f}^{(3)}_{k,k_{3},k_{13},k_{123}}\notag\\& 
+(k+k_{3})^{\mu_3} (k_{3}+k_{23})^{\mu_2} (k_{23}+k_{123})^{\mu_1}\mathrm{f}^{(3)}_{k,k_{3},k_{23},k_{123}}\notag\\& 
+2g^{\mu_1\mu_2} (k_{12}+k_{123})^{\mu_3}\mathrm{f}^{(2)}_{k,k_{12},k_{123}}\notag\\& 
+2g^{\mu_1\mu_3} (k_{13}+k_{123})^{\mu_2}\mathrm{f}^{(2)}_{k,k_{13},k_{123}}\notag\\& 
+2g^{\mu_2\mu_3} (k_{23}+k_{123})^{\mu_1}\mathrm{f}^{(2)}_{k,k_{23},k_{123}}\notag\\& 
+2g^{\mu_2\mu_3} (k+k_{1})^{\mu_1} \mathrm{f}^{(2)}_{k,k_{1},k_{123}}\notag\\& 
+2g^{\mu_1\mu_3} (k+k_{2})^{\mu_2} \mathrm{f}^{(2)}_{k,k_{2},k_{123}}\notag\\& 
+2g^{\mu_1\mu_2} (k+k_{3})^{\mu_3} \mathrm{f}^{(2)}_{k,k_{3},k_{123}}
\end{align}
additional momenta defined as $k_{i}=k+q_i$, $k_{ij}=k+q_i+q_j$,
$k_{ijk}=k+q_i+q_j+q_k$ and
the shorthand notations for first, second and third order finite-differences are introduced\footnote{ 
Similar abbreviations are used for finite differences of the mass function $m\rightarrow \mathrm{m}^{(i)}$
.}
\begin{align}
&\mathrm{f}^{(1)}_{p,q}     =\frac{f\left(  p\right)
		-f\left(  q\right)  }{p^{2}-q^{2}},\quad
\mathrm{f}^{(2)}_{p,q,l}     =\frac{\mathrm{f}^{(1)}_{p,q}  -\mathrm{f}^{(1)}_{p,l}}{q^2-l^2},\quad
\mathrm{f}^{(3)}_{p,q,l,k}     =\frac{\mathrm{f}^{(2)}_{p,q,l}  -\mathrm{f}^{(2)}_{p,q,k}}{l^2-k^2}
.\nonumber
\end{align}

\section{Cross-check of dressing and local limits}
\label{Crosscheck}
\begin{figure}[t]
\centering
\includegraphics[width=0.49\textwidth]{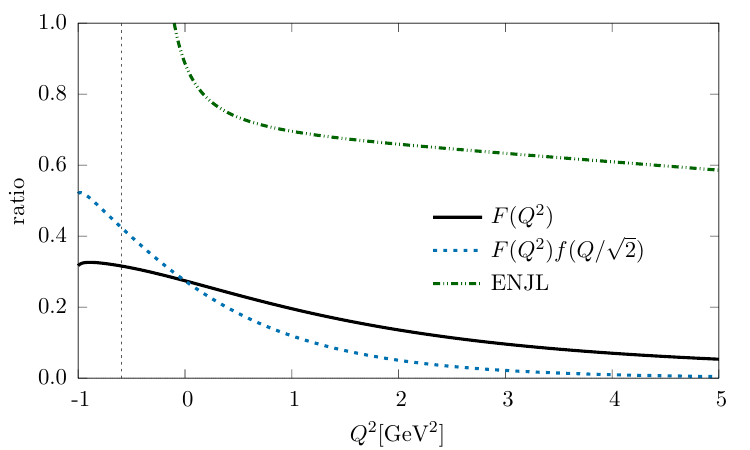}
\caption{Ratio of additional vertex with vector meson exchange with respect to VMD result $-\gamma^{\mu}_T{Q^2}/{(Q^2+M_V^2)}$. The green double-dashed line is the ENJL result \eqref{EqENJLAddVertex},  the black solid and blue dashed lines are  the functions $F(Q^2)$ and $F(Q^2)f(Q/\sqrt{2})$ in  the nonlocal model  \eqref{EqNonlocAddVertex}. The VMD result corresponds to 1.}
	\label{fig-RatioAdditionalVertex}       %
\end{figure}

The influence of photon dressing by vector mesons on LbL contributions is studied 
in \cite{Bijnens:1995cc} and \cite{Goecke:2012qm} 
within  the framework of the ENJL model and the DSE approach.
The effective vertex in the ENJL model \cite{Bijnens:1995cc} with vector meson exchange is\footnote{In ENJL model \cite{Bijnens:1995cc}, the momentum-dependent mass of the vector meson is expressed via using the incomplete Gamma function \cite{Bijnens:1994ey}
\begin{align}
M_V^2(Q^2)=\frac{N_c \Lambda_\chi^2}{8 \pi^2 G_V \bar{\Pi}^{(1)}(Q^2)} ,\quad \bar{\Pi}^{(1)}(Q^2) =\frac{N_c }{2 \pi^2 } \int_{0}^{1}dx\, x(1-x)\Gamma(0,x_{ij}) ,\quad x_{ij} =\frac{m_Q^2+Q^2 x(1-x)}{\Lambda_\chi^2}, \label{BijensMrho}
\end{align}
and model parameters \cite{Bijnens:1995cc} are $\Lambda_\chi=1.16$ GeV,  $G_V=1.263$ GeV$^{-2}$ and constituent quark mass is $m_Q=275$ MeV. At zero momentum, the momentum-dependent mass of the vector meson is $M_V^2(0)=0.678$ GeV$^2$ and for large $Q^2$ it is strongly increased, i.e. at $Q^2=4$ GeV$^2$ it is $M_V^2(4)\approx 3.55$ GeV$^2$.
} 

\begin{align}
\mathbf{\Gamma}^{\gamma;\mu}_{\mathrm{ENJL}}&={\Gamma}^{\gamma;\mu}_\mathrm{ENJL}+\Delta\mathbf{\Gamma}^{\gamma;\mu}_\mathrm{ENJL}=\gamma^{\mu}-\gamma^{\mu}_T\frac{Q^2}{Q^2+M_V^2(Q^2)},
\label{EqENJLAddVertex}
\end{align}
where $\gamma^{\mu}_T$ is the transversal projection of $\gamma^{\mu}$. The transversal component of the full vertex can be rewritten identically as
\begin{align}
\mathbf{\Gamma}^{\gamma;\mu}_{\mathrm{T,ENJL}}&\equiv \gamma^{\mu}_T\frac{M_V^2(Q^2)}{Q^2+M_V^2(Q^2)}.
\end{align}
In VMD approximation, which is discussed in \cite{Goecke:2012qm}, the vector meson mass is constant\footnote{In the DSE approach, the transition vertex can be approximated as follows \cite{Maris:1999bh,Goecke:2012qm}:
\begin{align}
\mathbf{\Gamma}^{\gamma;\mu}_{\mathrm{DSE}}&={\Gamma}^{\gamma;\mu}_\mathrm{BC}-\gamma^{\mu}_T\frac{\omega^4 N_V}{\omega^4+k^4}\frac{f_V}{M_V}\frac{Q^2}{Q^2+M_V^2}\mathrm{e}^{-\alpha(Q^2+M_V^2)},
\end{align}
where ${\Gamma}^{\gamma;\mu}_\mathrm{BC}$ is the vertex in Ball-Chiu ans{\"a}tze \cite{Ball:1980ay}.
}
, i.e. $M_V^2(Q^2)\equiv M_V^2$.
In the nonlocal model, the additional transition vertex with one photon \eqref{AdditionalDressedVertex}, can be effectively represented by
\begin{align}
\Delta\mathbf{\Gamma}^{\gamma;\mu}_{p_2,p_1,q_1}&=-\gamma^{\mu}_T \frac{Q^2}{Q^2+M_V^2} f(p_1) f(p_2) F(Q^2), \label{EqNonlocAddVertex}
\end{align}
where $F(Q^2)$ is a smooth function. For symmetric splitting, $p_{1,2}=k\pm Q/2$, ($k$ is the quark loop momentum) and a Gaussian form-factor, the combination of form-factors could be rewritten  $f(p_1) f(p_2)= f(\sqrt{2}k)f(Q/\sqrt{2})$ in an identical way.
In Fig. \ref{fig-RatioAdditionalVertex}, the behavior of the additional dressed function related to the VMD result is shown for the ENJL model and $F(Q^2), f(Q/\sqrt{2}) F(Q^2)$ for the nonlocal model, for a set with $m_d = 310$ MeV. One can see that the VMD, ENJL, and nonlocal models produce the same sign of the correction, and the result of the nonlocal model is suppressed compared to the VMD or ENJL models\footnote{It should be noted that constant for $\rho\to e^+e^-$ process $g_{\rho\gamma}=0.094$ GeV$^{2}$ deduced on $\rho$-meson mass-shell is not far from experimental value $g_{\rho\gamma}=0.121$ GeV$^{2}$.
}.

\begin{figure}[t]
\centering
\includegraphics[width=0.49\textwidth]{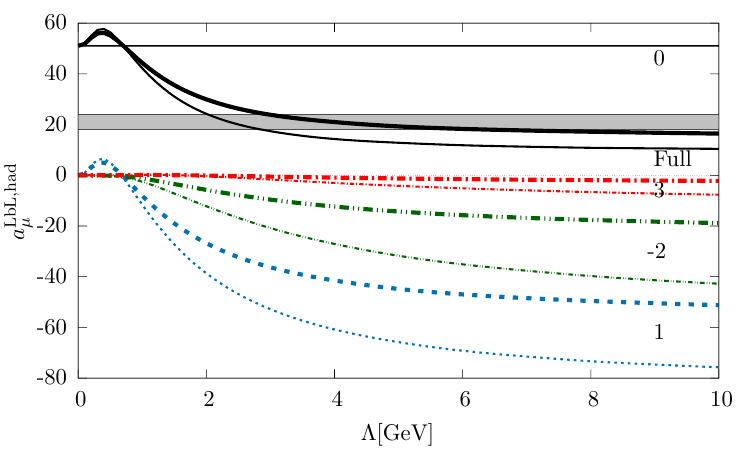}
\caption{(in units $10^{-11}$)
The LbL contribution to the muon AMM in the local limit for $m_Q= 300$ MeV.
The thin horizontal line marked as 0 is the quark loop contribution without vector mesons.
The contributions with vector mesons are: blue dashed line with 1 exchange, green dash-double-dotted line with 2 exchanges with minus sign and red dash-dotted is with 3 exchanges. The black solid line is the sum of all contributions: pure quark loop "0" plus vector meson exchanges "1+2+3". The thick lines correspond to the ENJL transition while the thin ones to the VMD one.
The shaded area represents the result of the ENJL model for the quark loop contribution \cite{Bijnens:1995xf}.
}
	\label{fig-VMDCheck}       %
\end{figure}

In order to cross-check the calculations, a case similar to that of ENJL or VMD could be considered. In this case, the value of the current quark mass is taken to be the constituent quark mass, $m_c\equiv m_Q = 300$ MeV as in \cite{Bijnens:1995xf}, the dynamical quark mass is set to zero $m_d=0$, the function $F(Q^2)$ in the transition loop \eqref{EqNonlocAddVertex} is set to $1$. %
In this case, one can study the behavior of the vector meson correction to the LbL contribution as a function of the $\Lambda$ parameter in the form factor of the vector meson vertex \eqref{EqENJLAddVertex}. The pure local result will be $\Lambda \to \infty$.
There are contributions with one, two and three vector meson exchanges, as already discussed. 
The overall factor for exchanges involving an odd number of vector mesons is negative compared to the contribution without vector mesons, while for even exchanges it is positive.
The behavior of the contributions with different numbers of vector meson exchanges as a function of the $\Lambda$ parameter is shown in Fig. \ref{fig-VMDCheck} for VMD and ENJL vector meson dressings. %
In the ENJL model, the mass of the vector mesons depends on the momentum \eqref{BijensMrho}. 
For infinitely large $\Lambda$, the total answers for VMD and ENJL are $9.4$ and $15.4$.
It should be noted that in \cite{Bijnens:1995xf}, the scheme for separating low and high energies is used with a scale parameter of $\mu \sim 0.7 - 4$ GeV. In this scheme, the low energy contribution is restricted such that all internal momentum is $|Q|\leq\mu$ in the integration procedure. For high energies, the contribution of a quark loop with mass $m_Q=\mu$ is taken. This separation is applied to all diagrams. 
One can see that the number $21$, for the ENJL result quoted in \cite{Bijnens:1995xf} roughly corresponds to $\Lambda \sim 4$ GeV.
On the other hand, in Fig. \ref{fig-VMDCheck}, one can see that even in the limit of constant mass, the value of the vector meson correction will be small if $\Lambda \sim 1$ GeV.

\begin{figure}[t]
\centering
\includegraphics[width=0.49\textwidth]{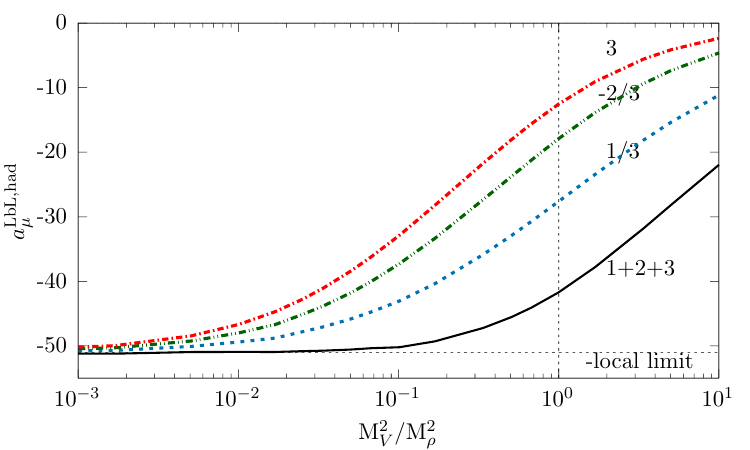}
\caption{(in units $10^{-11}$)
The LbL contribution from diagrams with additional vector meson vertices to the muon AMM in the local limit for $m_Q= 300$ MeV with VMD transition, as a function of the ratio of $M_V^2$ to the physical $M_\rho^2$.
The thin horizontal dashed line marked as the local limit is minus the quark loop contribution without vector mesons.
Thin vertical dashed line corresponds to physical point $M_V^2/M_\rho^2=1$.
The contributions with vector mesons are: blue dashed line with 1 exchanges (divided by factor 3), green double dash-dotted line with 2 exchanges with minus sign (divided by factor 3) and red dash-dotted line is with 3 exchanges. The black solid line is the sum of all vector-meson contributions. 
 }
	\label{fig-VMDCheckScale}       %
\end{figure}

On the other hand, in the local limit of $\Lambda \to \infty$, supplemented by the limit of zero vector-meson mass, the correction in equation \eqref{EqENJLAddVertex} becomes identical to $\Delta \mathbf{\Gamma}^{\gamma; \mu} = -\gamma_T^{\mu}$.
Therefore, in this limit, the contribution of diagrams involving vector meson exchanges becomes proportional to that of constituent quark loops. Since there are three diagrams for one vector meson exchange, three for two vector meson exchanges, and only one for three vector meson exchanges, the corresponding coefficients of proportionality are $-3$, $3$ and $-1$. The sum of all vector meson exchanges should equal the contribution of the constituent quark loop, with a minus sign. The behavior of the vector meson contributions is shown in Fig. \ref{fig-VMDCheckScale} as a function of its mass.

%

\end{document}